\documentclass[11pt]{article}
\usepackage{xspace}
\usepackage{graphicx}
\usepackage{amsmath}
\usepackage{amssymb}
\usepackage{color}
\usepackage{url}
\definecolor{Red}{rgb}{1,0,0}
\definecolor{Green}{rgb}{0,1,0}
\definecolor{Blue}{rgb}{0,0,1}
\definecolor{Black}{rgb}{0,0,0}

\def\beq{\begin{equation}}
\def\eeq#1{\label{#1}\end{equation}}
\def\eeqn{\end{equation}}

\def\beqa{\begin{eqnarray}}
\def\eeqa#1{\label{#1}\end{eqnarray}}
\def\eeqan{\end{eqnarray}}

\let\bar=\overbar

\def\Dslash{\not{\hbox{\kern-4pt $D$}}}
\def\dslash{\not{\hbox{\kern-2pt $\del$}}}

\def\msb{{\bar{\ssstyle M \kern -1pt S}}}

\def\Title#1{\begin{center} {\Large {\bf #1} } \end{center}}

\begin{document}

\Title{Neutrino Interactions and Long Baseline Physics}

\bigskip\bigskip


\begin{raggedright}

{\it Ulrich Mosel\index{Mosel, U.},\\
Institut fuer Theoretische Physik\\
Universitaet Giessen\\
D-35392 Giessen, Germany}\\

\end{raggedright}
\vspace{1.cm}

{\small
\begin{flushleft}
\emph{To appear in the proceedings of the Prospects in Neutrino Physics Conference, 15 -- 17 December, 2014, held at Queen Mary University of London, UK.}
\end{flushleft}
}

\section{Introduction}
The interactions of neutrinos with \emph{nucleons} can provide valuable information on axial properties and transition form factors \cite{Formaggio:2013kya}.  For example, the nucleon's axial form factor is rather badly known. It is usually reduced to a dipole ansatz, with one free parameter, the axial mass, remaining. This axial mass has been determined in many neutrino experiments on nucleons (or deuterons) and assumes a value of $M_A \approx 1$ GeV \cite{Bernard:2001rs}. The dipole form, however, cannot really be constrained further by experiment \cite{Bhattacharya:2011ah} and indeed the vector form factors obtained from elastic electron scattering show a significantly more complicated dependence on the squared four-momentum $Q^2$ \cite{Arrington:2006zm}. A similar situation exists for the transition form factors where even less is known. For example, for the $\Delta$ resonance the transition current involves 3 vector form factors and 3 axial ones. While the 3 vector form factors are reasonably well determined by elektron-induced pion production on the nucleon, the 3 axial form factors are largely unknown. Present data \cite{Formaggio:2013kya} seem to be sensitive to only one of them  and for it again the simple dipole form has been assumed.

The investigation of interactions of neutrinos with \emph{nuclei} may thus seem to be of only theoretical interest, since many elementary processes are not well understood. Indeed, there is a large theoretical interest in calculating the response of nuclei to external probes. This process can give valuable information about the nuclear many-body problem and the relevant reaction mechanisms, such as, e.g., the presence of interactions of the incoming probe with more than one nucleon \cite{Martini:2009uj,Martini:2010ex}. Such  processes require theoretical methods that go beyond the so-called impulse approximation in which the interaction proceeds just with one nucleon at a time. It is encouraging to see that these investigations have made a huge progress over the last 20 years. First, the impulse approximation was extended to work not only with quasi-free nucleons, but instead with dressed nucleons that contain some effects of interactions with the surrounding nucleons in the nucleus \cite{Benhar:1994hw}. These interactions can be partly summed up in 'spectral functions' which are given by the imaginary part of the nucleon's propagator inside the nucleus. The spectral functions contain effects of long-range mean-field potentials as well as effects of short-range correlations. Spectral functions have been very successfully employed in the description of the quasielastic response of nuclei to electrons \cite{Benhar:2006wy}. Not all of the many-body interactions can, however, be absorbed into a spectral function. The remaining interactions with neighboring nucleons then have to be treated explicitly by evaluating the many-body response through so-called 2 particle - 2 hole (2p2h) interactions \cite{Martini:2009uj,Martini:2010ex,Nieves:2011pp,Nieves:2011yp,Megias:2014qva}. Particularly noteworthy are recent ab-initio calculations of the quasielastic response of nuclei to incoming neutrinos that involves both vector and axial couplings \cite{Lovato:2014eva}. Due to their huge computational requirements such calculations have become available only during the last few years; they are so far restricted to non-relativistic energies and to rather low energy transfers. At higher energies and energy transfers also inelastic excitations, either through nucleon resonance excitations or through deep inelastic scattering (DIS), take place and the treatment must be relativistic.

While the investigation of interactions of neutrinos with \emph{nuclei} may seem to be of only theoretical interest there is a very practical interest in such studies for long-baseline experiments that look for neutrino oscillations, such as, e.g., T2K, MINOS, NOvA and, in the future DUNE (formerly called LBNE). In such experiments the neutrino flux at a far detector is compared with that at a near detector. From that comparison the neutrino oscillation parameters, mixing angles and a possibly CP-invariance violating phase, can be extracted. What is actually compared is the event rate (flux times cross section at a given neutrino energy $E_\nu$) at a far detector with that at the near detector. The flux comparison thus requires the knowledge of the neutrino energy. The complication lies in the fact that the neutrino energy is not known because of the special production method of neutrinos as secondary decay products of particles, mostly pions and kaons, that were produced in primary reactions of protons with nuclei. The neutrino energy thus must be reconstructed event by event from the final state of the reaction. Two methods for this reconstruction are being considered:

\begin{enumerate}
\item
The first method is a so-called calorimetric method in which the energy of the final state particles is observed. If the detector were perfect this would give directly the incoming beam energy, through energy conservation. Real-life detectors, however, have limitations. They have acceptance thresholds and problems to observe certain particle classes, e.g. electrically neutral particles. The experiment then sees actually only a small part of the energy of the final state phase-space and must extrapolate from that to the full final state energy. For that extrapolation so-called neutrino generators have been used. Necessary theoretical input here is the knowledge of the initial neutrino-nucleon interaction and the hadron-hadron interactions in the final state.

\item
The second method is based on the fact that for quasielastic charged current scattering of a neutrino on a free neutron at rest the incoming neutrino energy can be determined completely from the outgoing lepton kinematics (energy, angle). This method obviously requires an experimental veto on any other particles -- except for the proton -- in the final state, for a correct identification of the reaction mechanism as being quasielastic scattering. The complication now comes because all ongoing and planned long-baseline experiments use nuclear targets (C, O, Ar, Fe), partly in order to increase the reaction rates and partly because of experimental safety considerations. In real life the neutrons are thus not free, but they are bound in a nucleus and are Fermi-moving. Both of these facts lead to a smearing of the reconstructed energy around a sharp value, with an uncertainty width of about 60 MeV for a neutrino energy of about 1 GeV.

While this presents a natural lower limit to the error with which the neutrino energy can be reconstructed, there is a more difficult problem to overcome when using nuclear targets that leads to significantly larger errors. This is the correct identification of the scattering event as being quasielastic. While for a free nucleon all that is needed is an experimental veto on outgoing hadrons (besides the proton), in a nuclear target this is not enough, because here now pions that were originally produced can be absorbed. The final state can then not be distinguished from one after QE scattering. Indeed, the misidentification of QE scattering has led to the extraction of unphysical values for the axial mass from experiments with nuclear targes \cite{:2007ru,Adamson:2014pgc,Abe:2015oar}. This also means that pion production, either through nucleon resonances or DIS, is always entangled with QE in a wide sense (true QE on one nucleon and 2p2h processes). In other words: QE cross sections can never be determined by purely experimental means, but always require an event generator to subtract the pion contributions.
\end{enumerate}

\begin{figure}[!ht]
\begin{center}
\includegraphics[width=0.8\columnwidth]{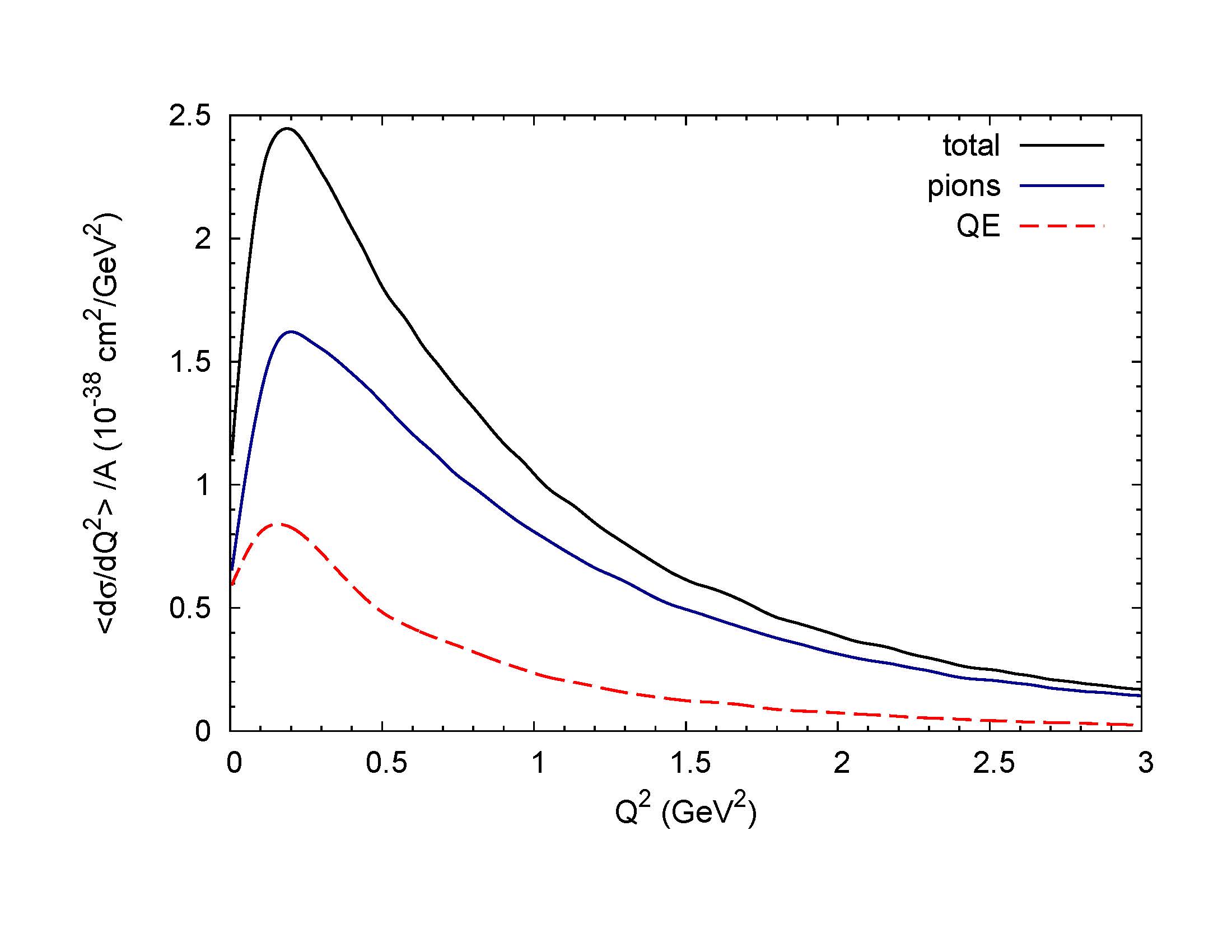}
\caption{(color online) Flux-averaged $Q^2$ distribution $\langle d\sigma/dQ^2 \rangle$ per nucleon for all events as a function of true $Q^2$ for an $^{40}Ar$ target in the LBNF beam. The contribution labeled 'pions' (solid blue line) gives the sum of all pion-producing processes (resonances, background and deep inelastic scattering), the one labeled 'QE' (dashed red) depicts the sum of true one-body CCQE and of 2p2h processes. The solid black line gives the sum of both (taken from~\cite{Mosel:2015oda}).}
\label{fig:QE-pi}
\end{center}
\end{figure}

 In order to get a feeling for the relative importance of QE scattering vs. pion production Fig.\ \ref{fig:QE-pi} shows these contributions for the flux expected at the near detector of the DUNE.
It is noticeable that pion production processes (resonances and DIS) contribute about 2/3 of the total cross sections and thus are the dominant component. Studies of the remainder, QE, then require first a quantitative understanding of pion production and an exact implementation of that understanding in generators. This, in turn, requires knowledge both of the neutrino-nucleon pion production process, of in-medium changes to that cross section for bound nucleons and a very good description of pion-nucleon interactions.

From these considerations it is clear that neutrino long-baseline experiments require a theoretical description of both the initial target ground state \emph{and} the full final state; just inclusive cross sections are not enough. These exeriments thus need input from nuclear structure and nuclear reaction theory. The former requires knowledge of the nuclear groundstate. The latter, on the other hand, requires knowledge of reaction mechanisms (1 particle vs. 2 particle initial processes) as well as a state-of-the-art description of final state interactions. While for the former the knowledge of lepton-nucleus interactions is essential, for the latter the essential requirement is knowledge of hadron-hadron interactions in the final state interactions. Without having these two areas quantitatively under control with state-of-the-art methods the goals of the precision era of neutrino physics can not be reached.

\section{Reaction mechanisms and oscillation signal}
We now have a closer look at the various reaction mechanisms contributing to the total neutrino-nucleus response in the LBNF beam: the nuclear target here is $^{40}$Ar. All results in the following have been obtained within the transport theoretical GiBUU framework \cite{Buss:2011mx,gibuu}. GiBUU goes beyond Monte Carlo simulations in that it takes the nuclear potentials into account. The code has been widely tested with the help of a broad range of nuclear reactions, with leptons, photons, hadrons and heavy ions as incoming beam.
\begin{figure}[!ht]
\begin{center}
\includegraphics[angle=0,width=0.8\columnwidth]{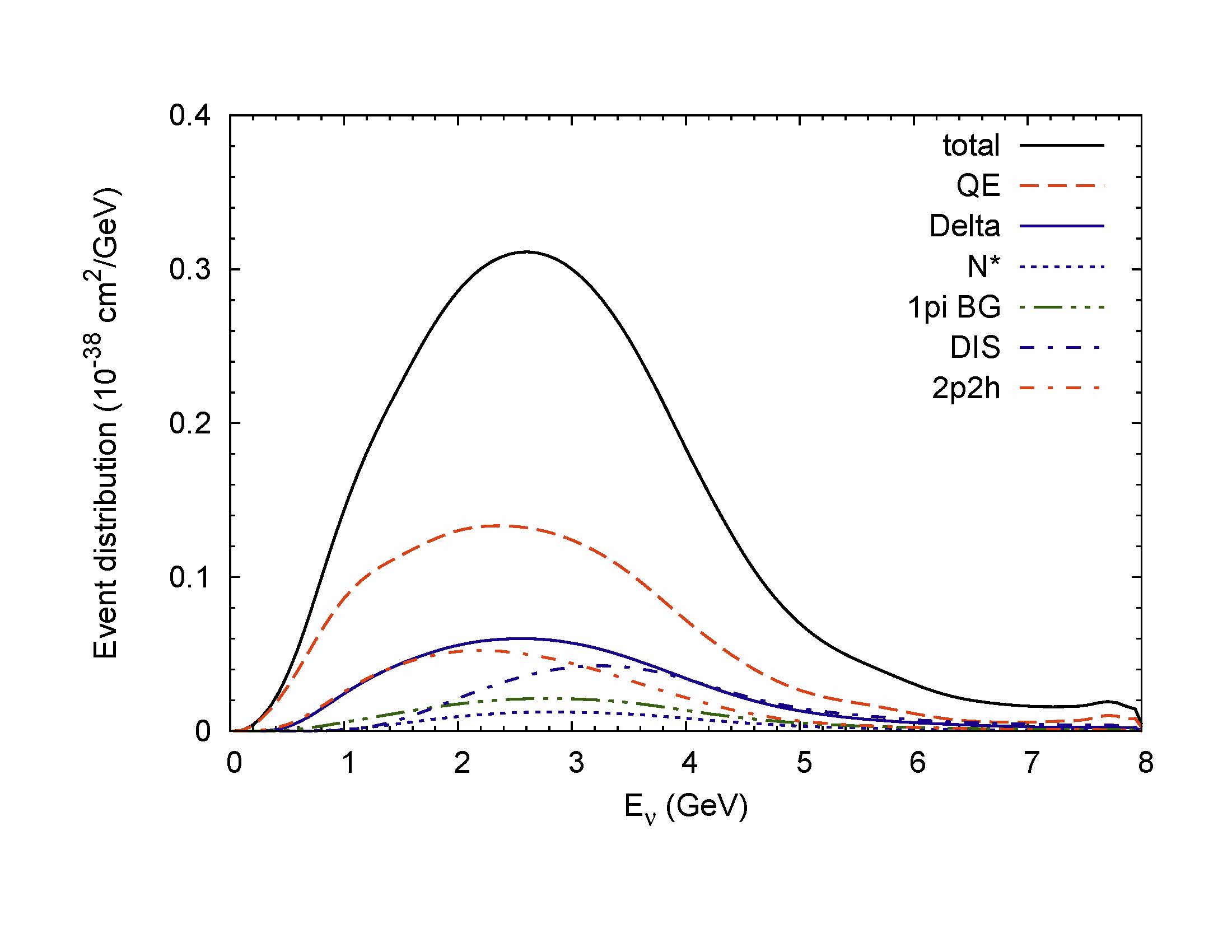}
\caption{(Color online) Event distribution (normalized flux times cross section) per nucleon for LBNE vs.\ true  energy (uppermost solid curve). Only events with 0 pions in the final state are taken into account. The various contributions to the total event rate are plotted as denoted in the figure (taken from~\cite{Mosel:2013fxa}).}\label{fig:events}
\end{center}
\end{figure}

Fig.\ \ref{fig:events}  shows the total event distribution and its decomposition into various production channels as a function of true neutrino energy. The true energies are those entering as input into the calculations. The overall energy-dependence of this event distribution is determined by the energy-distribution of the incoming neutrino beam. It is seen that -- at the peak -- true QE contributes about 1/3 of the total rate; the two next most important production channels are 2p-2h excitations and excitations of the $\Delta$ resonance. With a clear shift towards higher energies also DIS contributes on a similar level. The different processes thus exhibit different energy dependencies and they all overlap to a certain extent. This is a particular complication of experiments working with a neutrino flux in the few GeV region. At lower energies, e.g. the ones used by T2K, DIS and higher resonance excitations play no role so that essentially only $\Delta$ excitation contributes to pion production \cite{Lalakulich:2013iaa}. On the opposite, at significant higher beam energies larger than about 40 GeV the neutrino interactions are dominated by partonic degrees of freedom; there one has to deal with the complications of in-medium effects on partons (EMC effect). This figure also illustrates the difficulties one faces in determining cross sections from such measured event rates which are products of flux and cross section. Any error made in reconstructing the neutrino energy will directly translate into an error of the cross section.

\begin{figure}[!ht]
\begin{center}
\includegraphics[width=0.8\columnwidth]{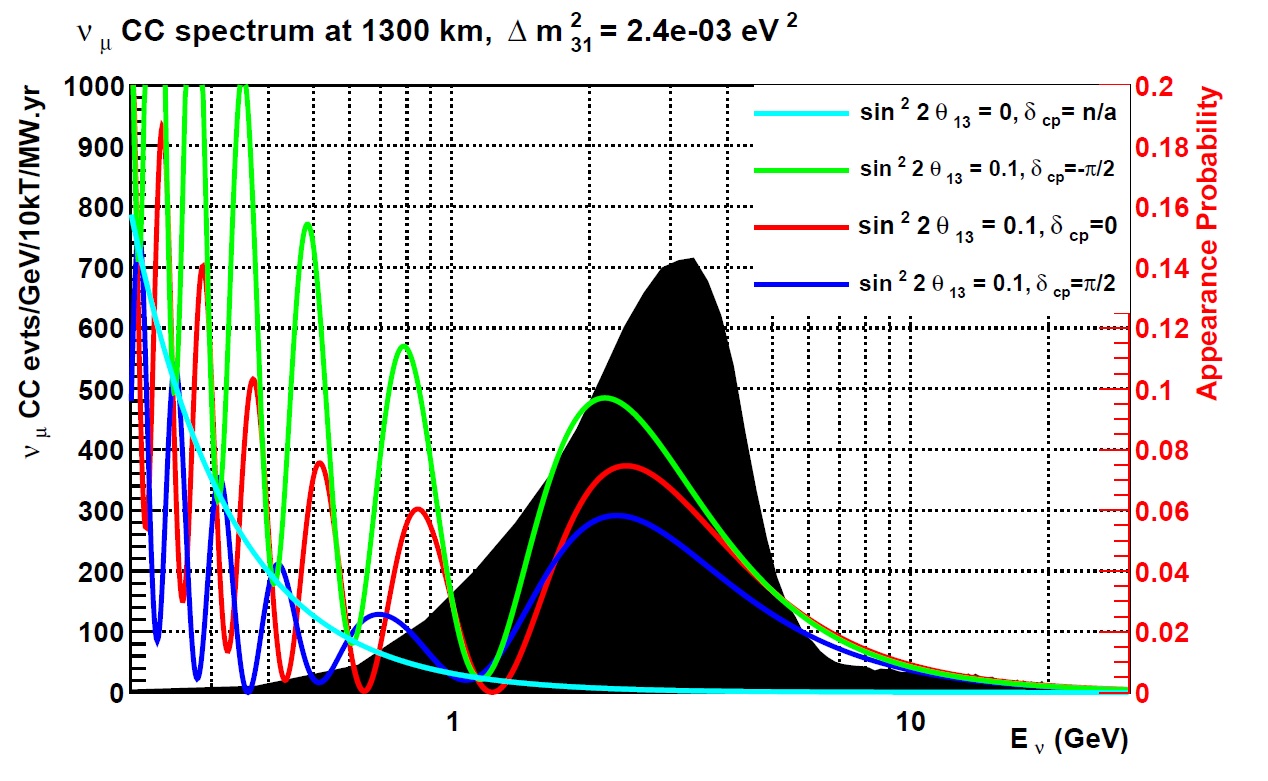}
\caption{$\nu_e$ appearance probability at a distance of 1300 km calculated for standard oscillation mixing angles. The four colored curves illustrate the sensitivity of the expected signal to the neutrino mixing angle $\theta_{13}$ and the CP-violating phase $\delta_{CP}$. The black peak shows the expected energy distribution for the neutrino beam  (taken from~\cite{Adams:2013qkq}).}
\label{fig:LBNE-oscill}
\end{center}
\end{figure}

In order to first get a feeling for the accuracy needed for the energy reconstruction in oscillation experiments a look at Fig. \ref{fig:LBNE-oscill} is helpful; this figure shows the expected oscillation signal for DUNE as a function as a function of neutrino Energy $E_\nu$ for some values of of two neutrino properties, mixing angle $\theta_{13}$ and the CP-violating phase $\delta_{CP}$.
The three curves under the flux profile can be distinguished from each other only if the neutrino energy can be determined to better than about 100 MeV. This gives a first hint at the accuracy that is needed at DUNE.
Errors in the energy reconstruction due to event-misidentification cause not just a shift of the energy axis, but instead distort the whole event distribution \cite{Lalakulich:2012hs,Martini:2012fa}.

\begin{figure}[!ht]
\begin{center}
\includegraphics[angle=0,width=0.7\columnwidth]{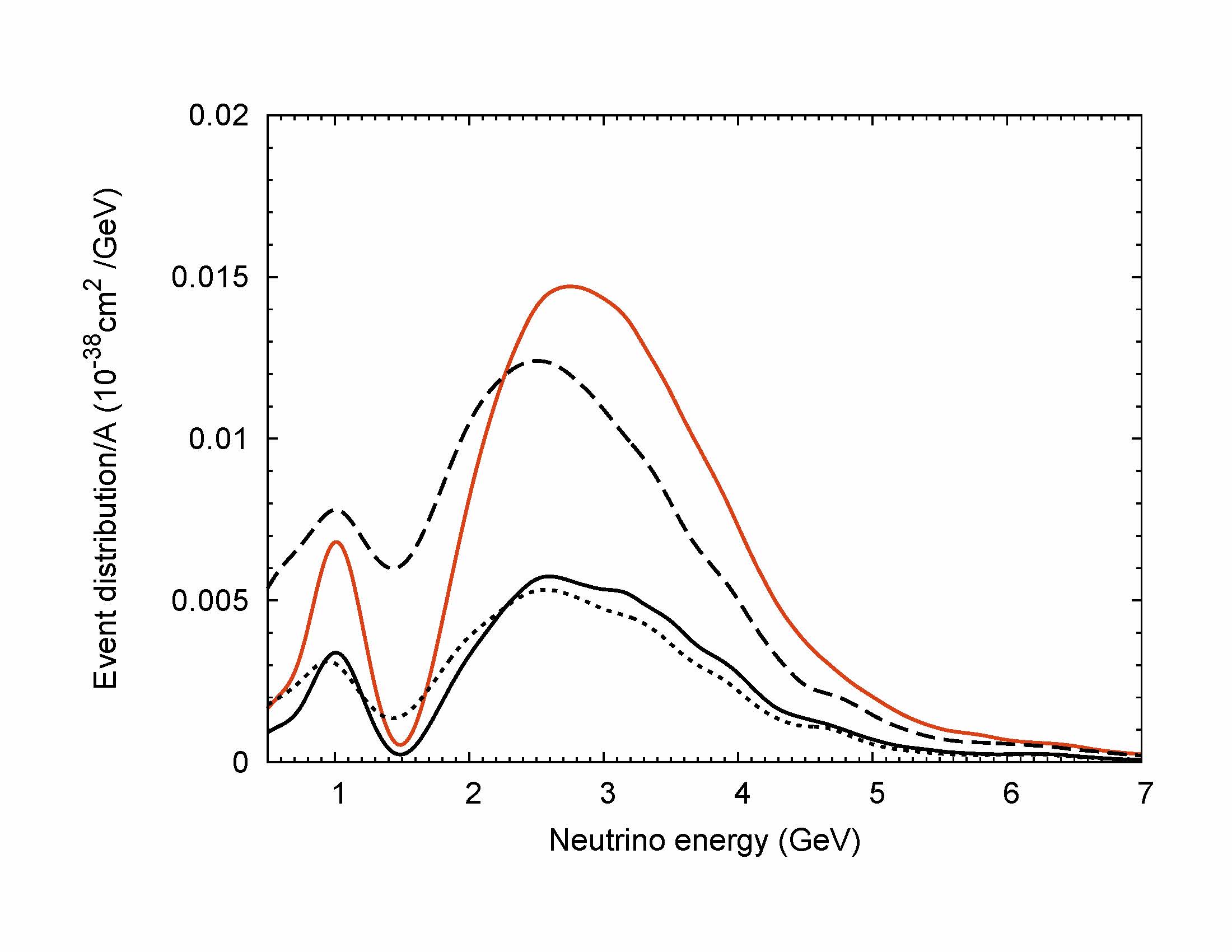}
\caption{$\nu_e$ appearance event distribution (normalized flux times cross section) per nucleon for LBNE vs.\ true (solid curve) and reconstructed (dashed curve) energy. The upper two curves show the results obtained from an event sample with 0 pions, the two lower curves are obtained from a sample with 0 pions, 1 proton and X neutrons (taken from~\cite{Mosel:2013fxa}).} \label{fig:mue-app}
\end{center}
\end{figure}

For the neutrino oscillation analysis the neutrino energy plays an essential role because it enters into the oscillation formula. Errors in the reconstructed energy thus have an impact on error in the mixing angles and phases. This is illustrated in Fig.\ \ref{fig:mue-app} which in its lower two curves shows the expected event distribution for electron appearance at the far detector in the LBNF beam. Here, the oscillation signal is plotted for two different event samples as a function of true neutrino energy (solid curves) as well as, on the other hand, of an energy reconstructed from the outgoing electron kinematics assuming a true QE process (dashed curves).
The upper two curves give the oscillation signal that is obtained from an event sample with 0 pions.
These two curves, calculated with GiBUU, directly correspond to the red curve in Fig.\ \ref{fig:LBNE-oscill}. The event rate vs.\ reconstructed energy (upper dashed curve) is distorted as compared to the one vs.\ true energy (upper solid curve) and shifted by more than 500 MeV in its maximum even though only events with 0 outgoing pions have been used for the reconstruction. This is clearly above the accuracy required to distinguish between the various parameter scenarios in Fig.\ \ref{fig:LBNE-oscill}.

 A drastic improvement happens when the event sample is further restricted to contain one and only one proton (plus any number of neutrons). Now the difference between the lower two curves is at most 100 MeV. Since the energy reconstruction is based on the dynamics of a true (1-body) process this implies that requiring 1 proton in addition to 0 pions gives a cleaner identification of true QE. This result depends crucially on the theoretical description of final state interactions, not just between pions and nuclei, but also between nucleons. The final state cascade must allow for an avalanche effect in which in most cases the outgoing proton from a neutrino-induced charged current scattering rescatters thus increasing the multiplicity of final state nucleons and and simultaneously lowering their energies. An event with only 1 final state proton then is a rather clean, direct signal of QE which has not undergone this final state scattering. This has recently also been exploited in a study of QE scattering by the MINER$\nu$A experiment \cite{Walton:2014esl}.

The same improvement also shows up for the difference between the true and the reconstructed oscillation signal in dependence on $\delta_{CP}$ \cite{Mosel:2013fxa}. Experiments looking for this phase would be well advised to look at events with 0 pions, 1 proton and X (unobserved) neutrons.

\section{Summary}
Neutrino oscillation parameters can be extracted only if the incoming neutrino energy is known. The latter has to be reconstructed from final state particles. This reconstruction requires knowledge of neutrino-nucleon interaction rates in medium and of final state interactions between the outgoing hadrons. It also requires a theory that is able to describe the complete time-development of the neutrino-nucleus reaction; just inclusive cross sections are not enough. Employing state-of-the-art theoretical methods of nuclear physics, both for the description of the ground state  \emph{and} of the reaction mechanisms, is essential to be able to take event generators out into new regions of energy and target mass. 

Over the last few years tremendous progress has been made in this respect. The theoretical methods are available and have also been -- at least partly -- implemented into theoretical descriptions of a neutrino-nucleus reaction. It has also been realized that presently used event generators often lag behind in their implementation of present-day's nuclear physics. However, it is clearly disconcerting for any nuclear theorist to see that even in very recent experiment analyses by leading experiments still outdated methods with unphysical parameters are being used. One of the two most glaring examples are the large, unphysical axial mas of $M_A \approx 1.25$ GeV that still underlies the description of QE scattering in recent experimental analyses as well as the still ongoing use of the Rein-Sehgal formfactors for nucleon resonances that are known to fail in their description of electroproduction data \cite{Graczyk:2007bc,Leitner:2008fg}. Both of these severe shortcomings have been realized now for about 6 years, but still have not caused any modification of the generators used by these experiments. This is a clearly very unsatisfactory situation, both from a nuclear theory point of view and from a neutrino experimental point of view, since it  runs counter to the ambitions of the precision era of neutrino physics that has now begun. How uncertainties in the generators used actually affect the oscillation mixing angles has been illustrated in recent work \cite{Coloma:2013rqa,Coloma:2013tba}.

\bigskip
\section{Acknowledgments}

I am grateful for the ongoing support by the whole GiBUU team.

This work has been supported by DFG.

\bibliographystyle{utphys}
\bibliography{nuclear}

\providecommand{\href}[2]{#2}\begingroup\raggedright\begin{thebibliography}{10}

\bibitem{Formaggio:2013kya}
J.~Formaggio and G.~Zeller, 
 \href{http://dx.doi.org/10.1103/RevModPhys.84.1307}{{\em
  Rev.Mod.Phys.} {\bfseries 84} (2012) 1307},
\href{http://arxiv.org/abs/1305.7513}{{\ttfamily arXiv:1305.7513 [hep-ex]}}.

\bibitem{Bernard:2001rs}
V.~Bernard, L.~Elouadrhiri, and U.~Meissner,
 \href{http://dx.doi.org/10.1088/0954-3899/28/1/201}{{\em
  J.Phys.G} {\bfseries G28} (2002) R1--R35},
\href{http://arxiv.org/abs/hep-ph/0107088}{{\ttfamily arXiv:hep-ph/0107088
  [hep-ph]}}.

\bibitem{Bhattacharya:2011ah}
B.~Bhattacharya, R.~J. Hill, and G.~Paz, 
  \href{http://dx.doi.org/10.1103/PhysRevD.84.073006}{{\em Phys.Rev.D}
  {\bfseries 84} (2011) 073006},
\href{http://arxiv.org/abs/1108.0423}{{\ttfamily arXiv:1108.0423 [hep-ph]}}.

\bibitem{Arrington:2006zm}
J.~Arrington, C.~Roberts, and J.~Zanotti, 
'' \href{http://dx.doi.org/10.1088/0954-3899/34/7/S03}{{\em
  J.Phys.G} {\bfseries G34} (2007) S23--S52},
\href{http://arxiv.org/abs/nucl-th/0611050}{{\ttfamily arXiv:nucl-th/0611050
  [nucl-th]}}.

\bibitem{Martini:2009uj}
M.~Martini, M.~Ericson, G.~Chanfray, and J.~Marteau, 
  \href{http://dx.doi.org/10.1103/PhysRevC.80.065501}{{\em Phys.Rev.C}
  {\bfseries 80} (2009) 065501},
\href{http://arxiv.org/abs/0910.2622}{{\ttfamily arXiv:0910.2622 [nucl-th]}}.

\bibitem{Martini:2010ex}
M.~Martini, M.~Ericson, G.~Chanfray, and J.~Marteau,
  \href{http://dx.doi.org/10.1103/PhysRevC.81.045502}{{\em Phys.Rev.C}
  {\bfseries 81} (2010) 045502},
\href{http://arxiv.org/abs/1002.4538}{{\ttfamily arXiv:1002.4538 [hep-ph]}}.

\bibitem{Benhar:1994hw}
O.~Benhar, A.~Fabrocini, S.~Fantoni, and I.~Sick, 
\href{http://dx.doi.org/10.1016/0375-9474(94)90920-2}{{\em Nucl.Phys.}
  {\bfseries A579} (1994) 493--517}.

\bibitem{Benhar:2006wy}
O.~Benhar, D.~Day, and I.~Sick, 
   \href{http://dx.doi.org/10.1103/RevModPhys.80.189}{{\em Rev.
  Mod. Phys.} {\bfseries 80} (2008) 189--224},
\href{http://arxiv.org/abs/nucl-ex/0603029}{{\ttfamily arXiv:nucl-ex/0603029}}.

\bibitem{Nieves:2011pp}
J.~Nieves, I.~Ruiz~Simo, and M.~Vicente~Vacas, 
  \href{http://dx.doi.org/10.1103/PhysRevC.83.045501}{{\em Phys.Rev.C}
  {\bfseries 83} (2011) 045501},
\href{http://arxiv.org/abs/1102.2777}{{\ttfamily arXiv:1102.2777 [hep-ph]}}.

\bibitem{Nieves:2011yp}
J.~Nieves, I.~Simo, and M.~Vacas, 
  \href{http://dx.doi.org/10.1016/j.physletb.2011.11.061}{{\em Phys.Lett.}
  {\bfseries B707} (2012) 72--75},
\href{http://arxiv.org/abs/1106.5374}{{\ttfamily arXiv:1106.5374 [hep-ph]}}.

\bibitem{Megias:2014qva}
G.~Megias, T.~Donnelly, O.~Moreno, C.~Williamson, J.~Caballero, {\em et~al.},
  \href{http://dx.doi.org/10.1103/PhysRevD.91.073004}{{\em Phys.Rev.}
  {\bfseries D91} no.~7, (2015) 073004},
\href{http://arxiv.org/abs/1412.1822}{{\ttfamily arXiv:1412.1822 [nucl-th]}}.

\bibitem{Lovato:2014eva}
A.~Lovato, S.~Gandolfi, J.~Carlson, S.~C. Pieper, and R.~Schiavilla,
  \href{http://dx.doi.org/10.1103/PhysRevLett.112.182502}{{\em Phys.Rev.Lett.}
  {\bfseries 112} no.~18, (2014) 182502},
\href{http://arxiv.org/abs/1401.2605}{{\ttfamily arXiv:1401.2605 [nucl-th]}}.

\bibitem{:2007ru}
{\bfseries MiniBooNE Collaboration} Collaboration, A.~Aguilar-Arevalo {\em
  et~al.}, 
 \href{http://dx.doi.org/10.1103/PhysRevLett.100.032301}{{\em
  Phys.Rev.Lett.} {\bfseries 100} (2008) 032301},
\href{http://arxiv.org/abs/0706.0926}{{\ttfamily arXiv:0706.0926 [hep-ex]}}.

\bibitem{Adamson:2014pgc}
{\bfseries MINOS} Collaboration, P.~Adamson {\em et~al.}, 
  \href{http://dx.doi.org/10.1103/PhysRevD.91.012005}{{\em Phys.Rev.}
  {\bfseries D91} no.~1, (2015) 012005},
\href{http://arxiv.org/abs/1410.8613}{{\ttfamily arXiv:1410.8613 [hep-ex]}}.

\bibitem{Abe:2015oar}
{\bfseries T2K} Collaboration, K.~Abe {\em et~al.}, 
\href{http://arxiv.org/abs/1503.07452}{{\ttfamily arXiv:1503.07452 [hep-ex]}}.

\bibitem{Mosel:2015oda}
U.~Mosel, 
\href{http://arxiv.org/abs/1501.03160}{{\ttfamily arXiv:1501.03160 [hep-ex]}}.

\bibitem{Buss:2011mx}
O.~Buss, T.~Gaitanos, K.~Gallmeister, H.~van Hees, M.~Kaskulov, {\em et~al.},
  \href{http://dx.doi.org/10.1016/j.physrep.2011.12.001}{{\em Phys.Rept.}
  {\bfseries 512} (2012) 1--124},
\href{http://arxiv.org/abs/1106.1344}{{\ttfamily arXiv:1106.1344 [hep-ph]}}.

\bibitem{gibuu}
Code available from:
\newblock \url{http://gibuu.hepforge.org}.

\bibitem{Mosel:2013fxa}
U.~Mosel, O.~Lalakulich, and K.~Gallmeister, 
  \href{http://dx.doi.org/10.1103/PhysRevLett.112.151802}{{\em Phys.Rev.Lett.}
  {\bfseries 112} (2014) 151802},
\href{http://arxiv.org/abs/1311.7288}{{\ttfamily arXiv:1311.7288 [nucl-th]}}.

\bibitem{Lalakulich:2013iaa}
O.~Lalakulich and U.~Mosel, 
  \href{http://dx.doi.org/10.1103/PhysRevC.88.017601}{{\em Phys.Rev.}
  {\bfseries C88} (2013) 017601},
\href{http://arxiv.org/abs/1305.3861}{{\ttfamily arXiv:1305.3861 [nucl-th]}}.

\bibitem{Adams:2013qkq}
{\bfseries LBNE Collaboration} Collaboration, C.~Adams {\em et~al.},
\href{http://arxiv.org/abs/1307.7335}{{\ttfamily arXiv:1307.7335 [hep-ex]}}.

\bibitem{Lalakulich:2012hs}
O.~Lalakulich, U.~Mosel, and K.~Gallmeister, 
  \href{http://dx.doi.org/10.1103/PhysRevC.86.054606}{{\em Phys.Rev.}
  {\bfseries C86} (2012) 054606},
\href{http://arxiv.org/abs/1208.3678}{{\ttfamily arXiv:1208.3678 [nucl-th]}}.

\bibitem{Martini:2012fa}
M.~Martini, M.~Ericson, and G.~Chanfray, 
  \href{http://dx.doi.org/10.1103/PhysRevD.85.093012}{{\em Phys.Rev.}
  {\bfseries D85} (2012) 093012},
\href{http://arxiv.org/abs/1202.4745}{{\ttfamily arXiv:1202.4745 [hep-ph]}}.

\bibitem{Walton:2014esl}
{\bfseries MINERvA Collaboration} Collaboration, T.~Walton {\em et~al.},
\href{http://arxiv.org/abs/1409.4497}{{\ttfamily arXiv:1409.4497 [hep-ex]}}.

\bibitem{Graczyk:2007bc}
K.~M. Graczyk and J.~T. Sobczyk, 
\href{http://dx.doi.org/10.1103/PhysRevD.77.053001}{{\em Phys.
  Rev.} {\bfseries D77} (2008) 053001},
\href{http://arxiv.org/abs/0707.3561}{{\ttfamily arXiv:0707.3561 [hep-ph]}}.

\bibitem{Leitner:2008fg}
T.~Leitner, O.~Buss, U.~Mosel, and L.~Alvarez-Ruso, 
{\em PoS} {\bfseries NUFACT08} (2008) 009,
\href{http://arxiv.org/abs/0809.3986}{{\ttfamily arXiv:0809.3986 [nucl-th]}}.

\bibitem{Coloma:2013rqa}
P.~Coloma and P.~Huber,
\href{http://dx.doi.org/abs/10.1103/PhysRevLett.111.221802}{{\em Phys.Rev.Lett.}
{\bfseries 22} (2013) 221802},
\href{http://arxiv.org/abs/1307.1243}{{\ttfamily arXiv:1307.1243 [hep-ph]}}.

\bibitem{Coloma:2013tba}
P.~Coloma, P.~Huber, C.-M. Jen, and C.~Mariani, 
  \href{http://dx.doi.org/10.1103/PhysRevD.89.073015}{{\em Phys.Rev.}
  {\bfseries D89} (2014) 073015},
\href{http://arxiv.org/abs/1311.4506}{{\ttfamily arXiv:1311.4506 [hep-ph]}}.

\end{thebibliography}\endgroup

\end{document}